\documentclass[aps,twocolumn,pra,tightenlines,floatfix,showpacs]{revtex4}

\usepackage[dvips]{graphicx}
\usepackage[english]{babel}
\usepackage{amsmath}
\usepackage{amssymb}
\usepackage{slashed}

\usepackage[dvips]{graphicx}
\newcommand{\be}{\begin{equation}}
\newcommand{\ee}{\end{equation}}
\newcommand{\ba}{\begin{eqnarray}}
\newcommand{\ea}{\end{eqnarray}}

\newcommand{\Ek}{E_{\mathbf{k}}}

\newcommand{\xik}{\xi_{\mathbf{k}}}
\newcommand{\sumk}{\sum_{\mathbf{k}}}

\newcommand{\sumq}{\sum_{\mathbf{q}}}

\newcommand{\p}{\partial}

\begin{document}

\title{The Compressibility in Strongly Correlated
Superconductors and Superfluids: From BCS to BEC}

\author{Hao Guo$^{1}$, Yan He$^2$, Chih-Chun Chien$^3$, K. Levin$^2$}
\affiliation{$^1$Department of Physics, Southeast University, Nanjing 211189, China}
\affiliation{$^2$James Franck Institute and Department of Physics,
University of Chicago, Chicago, Illinois 60637, USA}
\affiliation{$^3$Theoretical Division, Los Alamos National Laboratory, Los Alamos, NM, 87545, USA}

\date{\today}

\pacs{03.75.Ss,74.20.Fg,67.85.-d}

\begin{abstract}
We present a theoretical study of the compressibility, $\kappa$, in a Fermi gas with attractive contact interactions,
providing predictions for the strongly-attractive regime and the superfluid
phase.
Our work emphasizes
the compressibility sum rule and gauge invariance as constraints on
$\kappa$ and we show
how within a particular $t$-matrix approach,
these can be satisfied in the normal phase when no approximations are made.
For tractability, approximations must be introduced,
and it is believed that thermodynamical approaches to $\kappa$
are more reliable,
than correlation function based schemes. Contrasting
with
other studies in the literature, we present
thermodynamic calculations of $\kappa$; these yield
semi-quantitative agreement with experiment and provide
physical insight into similar results obtained via quantum Monte Carlo
simulations.
\end{abstract}

\maketitle

There is extensive recent literature on
response functions in strongly correlated
superconductors and the counterpart atomic Fermi gas superfluids.
Here the correlations are presumed sufficiently strong so that
the classic BCS theory, which has been remarkably successful
for many decades, is no longer adequate.
Among the experiments of interest are
thermodynamic response functions \cite{Zwierlein12,Ketterle11}
as well as dynamical response studies \cite{ValePRL} in the Fermi
gases which undergo BCS-Bose-Einstein condensation (BEC) crossover.
Possibly related are novel probes of the density correlations in the
copper oxide superconductors \cite{Sawatzky13}.

A particularly important quantity derivable from the response function is the
compressibility, $\kappa$, because it provides direct
signatures of the transition temperature, $T_c$, which are
otherwise difficult to identify in
neutral superfluids.
Here we discuss the behavior of
$\kappa$, making contact with recent Fermi gas experiments \cite{Zwierlein12}.
Other groups \cite{StrinatiPRL12,Haussmann12} have computed $\kappa$, for
$T$ strictly above $T_c$.
To calibrate past and future work,
we discuss the
pitfalls \cite{Mahanbook} associated with calculating $\kappa$ in general many-body
theories.
A consistent theory of the response
functions must obey the appropriate ``Ward identity", which imposes gauge invariance
on the quantum correlation functions.
In this way different theoretical approaches can be assessed
according to whether they satisfy
the so-called longitudinal and transverse f-sum rules. Moreover,
implementation of this gauge invariance is particularly
complicated for $T < T_c$ where collective mode
effects enter into the density response \cite{OurPed}.

Added to this complication is the fact that
there are two distinct ways of arriving at a static
response such as the
compressibility: either through the zero frequency, zero momentum
limit
of the correlation function or via direct application
of thermodynamics. Consistency is equivalent to
imposing a sum rule known as the compressibility sum rule.
This latter constraint
is
known to be problematic in almost every approximate many body
theory from quantum hall liquids \cite{HalperinQHE} to the random-phase approximation (RPA) of
electron gases \cite{Singwi}.

Here we discuss a
``$Q$-limit Ward identity" \cite{Maebashi09},
which is equivalent to the compressibility sum rule
and show
how it is necessarily obeyed
in our microscopic theory of fermionic superfluids in BCS-BEC crossover.
However, for concrete calculations some approximations
are required. With these approximations we are able to
demonstrate consistency with gauge invariance through analytically
satisfying the
longitudinal and transverse f-sum rules, but are unable to
satisfy the compressibility sum rule.
Once approximations are made, as they have been in all BCS-BEC crossover calculations in the
literature \cite{StrinatiPRL12,Haussmann12}, this consistency requirement is forfeited.

The
compressibility
$\kappa \equiv n^{-2}(\partial n / \partial \mu)$
must be consistent with the compressibility sum rule \cite{Kubo_book}
\ba
\frac{\partial n}{\partial \mu}
&=&-K^{00}(\omega=0,\mathbf{q}\rightarrow\mathbf{0}),
\label{eq:3}
\ea
where $K^{00}$ is
the density-density component of response function tensor $K^{\mu\nu}$(Q).
 defined by
$J^{\mu}(Q) = K^{\mu \nu}(Q) A_{\nu}(Q)$, where the four-vector potential
$A^{\mu} = (\phi, \mathbf{A})$ incorporates the scalar and vector potential.
While our theory is applied to neutral superfluids, we contemplate
a weak fictitious electromagnetic (EM) field.

When the compressibility
constraint (or the equivalent sum rule) cannot be satisfied, as in the RPA of electron gases,
it has been suggested \cite{Singwi} that the
more meaningful answer is obtained via a thermodynamic route, where
gauge invariance, etc. plays a lesser role.
Here we present our results for $\kappa$ on either side of
unitarity obtained directly from thermodynamics,
where we can address the physics both above and below $T_c$.
We find that the compressibility increases as $T$ approaches $T_c$ from above, more
dramatically than in a Fermi liquid and that it never diverges, but rather
undergoes an upturned step discontinuity at the transition.
This upturn reflects the increasing bosonic character, as pairs are formed.
Below $T_c$, the behavior is
not so different from that in the BCS regime
where it reflects the
growth of a fermionic gap which tends to depress $\kappa$.
These results appear
in semi-quantitative agreement
with experiment \cite{Zwierlein12} in both the normal and superfluid phases.

%

The
gauge invariant EM response kernel can be expressed as
\begin{eqnarray}\label{KmnO}
K^{\mu\nu}(Q)&=&2\sum_{P}\Gamma^{\mu}(P+Q,P)G(P+Q)\times\nonumber\\
& &\gamma^{\nu}(P,P+Q)G(P)+\frac{n}{m}h^{\mu\nu},
\end{eqnarray}
where $h^{\mu\nu} \equiv-g^{\mu\nu}(1-g^{\nu0})$
and the diagonal metric tensor $g^{\mu \nu} $ is $(1,-1,-1,-1)$.
Throughout we define
$Q\equiv q^{\mu}=(i\Omega_{l},\mathbf{q})$ with $\Omega_{l}$ being the boson Matsubara frequency.
Similarly,
$P\equiv p^{\mu}=(i\omega_n,\mathbf{p})$ is the 4-momentum of the fermion with $\omega_n$ being the fermion Matsubara frequency.
The goal of linear response theory is to find the full EM vertex $\Gamma^{\mu}$ associated with the EM  response kernel                                                                        $ K ^{\mu\nu}(Q)$.
This full EM vertex must obey the Ward identity
$q_{\mu}\Gamma^{\mu}(P+Q,P)=G^{-1}(P+Q)-G^{-1}(P)$, which implies that
$q_{\mu}K^{\mu\nu}(Q)=0$  \cite{OurPed}.
Here we have introduced
the bare EM vertex
$\gamma^{\mu}(P+Q,P)=(1,\frac{\mathbf{p}+\frac{\mathbf{q}}{2}}{m})$.
The noninteracting Green's function is given by $G_0(P)=(i\omega_n-\xi_{\mathbf{p}})^{-1}$ with $\xi_{\mathbf{p}}=\frac{p^2}{2m}-\mu$.
 $G(P)$ is the single-particle Green's function determined by $G^{-1}(P)=G^{-1}_0(P)-\Sigma(P)$, where $\Sigma(P)$ is the fermion self-energy.
Different theories of BCS-BEC crossover will assume different forms for
$\Sigma(P)$.

Gauge invariance guarantees that the longitudinal and transverse
sum rules are satisfied.
A necessary and sufficient condition for the validity of the compressibility sum rule \eqref{eq:3} is the $Q$-limit Ward identity, which can be proven as follows.
We have
\begin{eqnarray}\label{CSR}
\frac{\partial n}{\partial \mu}&=&2\sum_P\frac{\partial G(P)}{\partial \mu}
=-2\sum_PG^2(P)\Big(1-\frac{\partial \Sigma(P)}{\partial \mu}\Big).
\end{eqnarray}
We show below that this leads to
\begin{equation}\label{QWI}
\Gamma^0(P,P)|_{\Omega=0,\mathbf{q}\rightarrow \mathbf{0}}
=
1-\frac{\partial \Sigma(P)}{\partial \mu},
\end{equation}
which we refer to as the
$Q$-limit Ward identity. Here $\Omega$ is the analytic continuation of $i\Omega_{l}$.
Indeed, comparing with the expression for $K^{00}(\omega=0,\mathbf{q}\rightarrow\mathbf{0})$ given by Eq.(\ref{KmnO}), we find
$\frac{\partial n}{\partial \mu}
=-2\sum_P\Gamma^0(P,P)G(P)\gamma^0(P,P)G(P)$.
When Eq.~(\ref{QWI}) is satisfied,
the compressibility obtained via thermodynamic
arguments is related to a
two-particle correlation function ($K^{00}$ in this case).
The $Q$-limit Ward identity serves as an independent constraint on linear response theories \cite{OurPed,order_note},
separate from
the Ward identity reflecting
gauge invariance. Because both make a connection
between the self energy and the vertex functions, they
pose severe challenges to a proper formulation of linear response theory.

We next demonstrate how these consistency conditions are
satisfied
in
BCS-BEC crossover in the pseudogap (pg) phase above $T_c$.  As in all analytic such schemes, we begin
with a $t$-matrix approach \cite{ourreview},
where the
propagator for the non-condensed pairs is generically given by
\begin{equation}
t_{\textrm{pg}}^{-1}(Q)=g^{-1}+\chi(Q)\,,
\label{eq:1}
\end{equation}
Here $g$ is the attractive coupling constant in the Hamiltonian
and $\chi$ is the pair susceptibility.  To capture the physics
of Gor'kov theory
\cite{ourreview} we
take
$\chi(Q)=\sum_{K}G_{0}(Q-K)G(K)$,
with
\begin{equation}
\label{eq:10}
\Sigma(K) \equiv \Sigma_{\textrm{pg}}(K)=\sum_{Q}t_{pg}(Q) G_{0}(Q-K).
\end{equation}

The diagrams which are consistent with particle number
conservation \cite{ourreview} consist of three types in addition to the bare vertex. They are the so-called Maki
Thompson (MT) contribution and two versions of the
Aslamazov-Larkin (AL) diagram. These have been presented in the
literature \cite{Kosztin2} and given by
\begin{eqnarray}
&&\textrm{MT}^{\mu}_{\textrm{pg}}(P+Q,P)=\sum_Kt_{\textrm{pg}}(K)G_0(K-P-Q)\nonumber\\
&&\quad\times G_0(K-P)\gamma^{\mu}(K-P,K-P-Q); \\
&&\textrm{AL}^{\mu}_1(P+Q,P)=-\sum_{K,L}t_{\textrm{pg}}(K)t_{\textrm{pg}}(K+Q)G_0(K-P)\nonumber\\
&&\quad\times G(L)G_0(K-L+Q)G_0(K-L)\nonumber\\
&&\quad\times\gamma^{\mu}(K-L+Q,K-L); \\
&&\textrm{AL}^{\mu}_2(P+Q,P)=-\sum_{K,L}t_{\textrm{pg}}(K)t_{\textrm{pg}}(K+Q)G_0(K-P)\nonumber\\
&&\quad\times G_0(K-L)G(L)G_0(L+Q)\Gamma^{\mu}(L+Q,L).
\end{eqnarray}
These diagrams are obtained
by inserting the EM vertex in the self-energy diagram in all possible ways and
can be shown to be fully consistent with the self energy
so that
\begin{eqnarray}
& &q_{\mu}\big[\textrm{MT}^{\mu}_{\textrm{pg}}(P+Q,P)+\textrm{AL}^{\mu}_1(P+Q,P)\nonumber\\
& &+\textrm{AL}^{\mu}_2(P+Q,P)\big]=\Sigma(P)-\Sigma(P+Q).
\end{eqnarray}
We write
\ba
-\frac{\p\Sigma_{\textrm{pg}}}{\p\mu}
=\textrm{MT}^0_{\textrm{pg}}(P,P)+\textrm{AL}^0_1(P,P)+\textrm{AL}^0_2(P,P).
\ea
Now if we combine the above results with the bare vertex $\gamma^{\mu}$, we find
that the Ward identity guaranteeing gauge invariance and
the Q-limit Ward Identity (\ref{QWI}) guaranteeing the compressibility sum rule are both satisfied, providing no
approximations \cite{Note2} are made.

While we have proved the compressibility sum rule on general
grounds, it is not in a particularly useful form
for numerical application.
To make things more transparent and tractable
we approximate the normal state
contribution to the self energy for temperatures
above but near $T_c$, where $t_{\textrm{pg}}(Q)$ is peaked near $Q = 0$.
We write
\begin{equation}
\Sigma_{\textrm{pg}}(K) = \sum_Q t_{\textrm{pg}}(Q) G_{0}(Q-K) \approx
G_{0} (-K)
\sum_Q t_{\textrm{pg}}(Q)
\,.
\label{eq:70}
\end{equation}
so that
\begin{equation}
\Sigma_{\textrm{pg}} (K)\approx -G_{0} (-K) \Delta_{\textrm{pg}}^2 \,,
\label{eq:sigma3}
\end{equation}
When extended to include the effects of
the condensate, the superconducting order parameter
$\Delta_{\textrm{sc}}$ is added to Eq.(\ref{eq:70}),
with the usual self energy
$\Sigma_{\textrm{sc}} = \frac {\Delta_{\textrm{sc}}^2}
{\omega + \xi_{\bf k}}$.
In this lowest order approximation
we see that the contributions to the self energy from
the condensed and non-condensed pairs are not distinguished.
The effective gap for fermionic excitations is
$\Delta = \sqrt{\Delta_{\textrm{sc}}^2 + \Delta_{\textrm{pg}}^2}$.
%
Above $T_c$, this self energy approximation
will lead to an expression for the density-density response function
discussed in detail elsewhere \cite{HaoPRL10}.
Importantly this result is analytically consistent with the longitudinal
and transverse f-sum rules.
\begin{figure*}
\centerline{\includegraphics[width=6.0in,clip]
{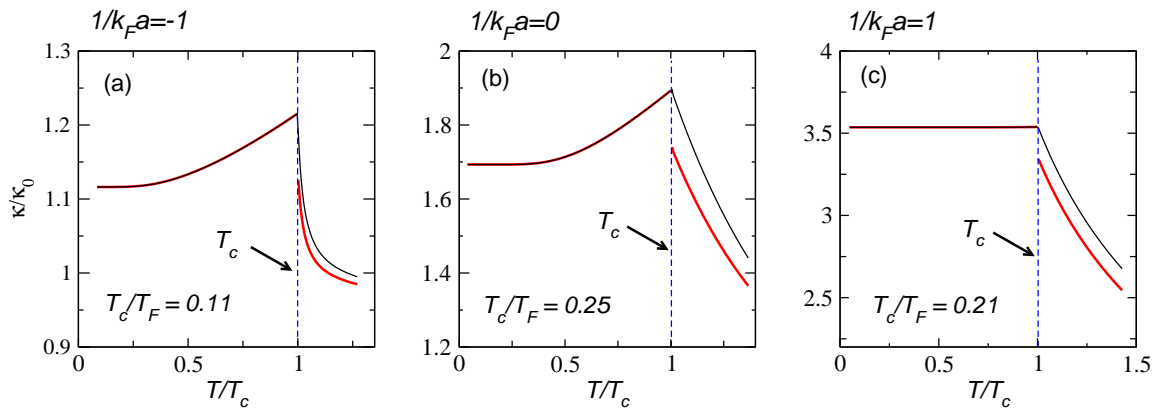}}
\caption{Our BCS-BEC crossover (thermodynamic approach) to
compressibility with finite $\gamma$ (red)
and $\gamma=0$ (black) on (a) BCS side, (b) unitarity, and (c) BEC side. The upturn in the normal state as $T$ decreases is due to the
bosonic contribution.
}
\label{fig:1}
\end{figure*}
As in general theories \cite{Mahanbook,Singwi,HalperinQHE}, not
surprisingly,
this approximate density-density response
function violates the compressibility sum rule.
This means that we will find two different answers for
the compressibility via thermodynamics and the response function.

It has been argued \cite{Singwi} that
the
thermodynamical approach
is more reliable.
This is due in part to the complexity of satisfying diagrammatic
consistency requirements. We, thus, turn to this thermodynamic-based approach \cite{ourlongpaper}, later
incorporating a parameter choice from radio-frequency spectroscopy \cite{RFlong}.
As one departs from the BCS regime,
there
are additional ``bosonic" contributions due to strong pairing fluctuations, besides those associated with the fermionic
excitations.
The resulting
thermodynamical potential
$\Omega=\Omega_f+\Omega_b$ including the fermions (f) and composite-bosons (b)
can be written as
\begin{eqnarray}
\Omega_f&=&-\frac{\Delta^2}{g}+\sumk[(\xik-\Ek)-\frac{2}{\beta}\ln(1+e^{-\beta\Ek})], \\
\Omega_b&=&-a_0\Delta^2\mu_p+\sumq \frac{1}{\beta}\ln(1-e^{-\beta\omega_q}).
\end{eqnarray}
Here $\mu_p$ is the pair chemical potential which is non-zero above
and zero below $T_c$ and
$\omega_q$ is the pair dispersion, which (along with the residue $a_0$)
arises from a small $Q$ expansion
of the t-matrix, $t_{pg}$. $\beta=(k_B T)^{-1}$ and we set $k_{B}\equiv 1$.
The coupling constant $g$ is related to the $s$-wave scattering length $a$ \cite{ourreview,RFlong}.
Note that $\Omega_f$ has a similar structure as that found in
BCS theory, but implicity involves composite-boson contributions.

This thermodynamic potential then yields
self consistent conditions on the gap, pseudo-gap and chemical potential via
variational conditions:
$\frac{\delta\Omega}{\delta\Delta}=0$, $\frac{\delta\Omega}{\delta\mu_p}=0$, and
$n=-\frac{\delta\Omega}{\delta\mu}$.
For example, these self consistent equations lead to the gap
equation
\begin{equation}
g^{-1} + \sum_k \frac {1 - 2 f(E_{\mathbf{k}}) }{2 E_{\mathbf{k}}}
= a_0 \mu_p,
\end{equation}
which is of the familiar BCS form for
$T < T_c$, and extends naturally to include a pairing gap in the
normal phase as well. One important feature of this thermodynamic approach is
that the superfluid transition is second order \cite{RFlong,OurAnnPhys}, in contrast to the artificial
first order transition found in other work
\cite{firstordertransitionpapers_full}.
The
compressibility is
\begin{eqnarray}
\frac{\p n}{\p\mu}=\Big(\frac{\p
n}{\p\mu}\Big)_{\Delta}+\Big(\frac{\p n}{\p\Delta}\Big)_{\mu}
\frac{\p\Delta}{\p\mu}
+
\Big(\frac{\p n}{\p\mu_p}\Big)_{\mu}
\frac{\p\mu_p}{\p\mu},
\end{eqnarray}
where this last term vanishes below $T_c$.
From the
number equation $n=-\frac{\delta\Omega}{\delta\mu}$,
we obtain the following expression for the compressibility \textit{above $T_c$},
$\big(\frac{\partial n}{\partial \mu}\big)_{T > T_c}$, given by
\begin{eqnarray}
I_1 +\frac{\Delta^2 \Big[\sum_{\mathbf{p}}
\frac{\xi_{\mathbf{p}}}{E^2_{\mathbf{p}}}\big(\frac{1-2f(E_{\mathbf{p}})}
{E_{\mathbf{p}}}
+2\frac{\partial f(E_{\mathbf{p}})}{\partial E_{\mathbf{p}}}\big)\Big]^2}
{\sum_{\mathbf{p}}\frac{\Delta^2}{E^2_{\mathbf{p}}} \big(\frac{1-2f(E_{\mathbf
{p}}
)}
{E_{\mathbf{p}}}+2\frac{\partial f(E_{\mathbf{p}})}{\partial E_{\mathbf{p}}}
\big)
-\frac{4 a_o^2 \Delta^2}{\sum_q b'(\omega_q)}
}.
\label{eq:18}
\end{eqnarray}
Here $b'(x)$ is the derivative of the usual Bose-Einstein
function with respect to its argument.
We
define
\begin{equation}
I_1 \equiv \sum_{\mathbf{p}}\Big[\frac{\Delta^2}{E^3_{\mathbf{p}}}\big(1-2f(E_{\mathbf{p}
})
\big)
-2\frac{\xi^2_{\mathbf{p}}}{E^2_{\mathbf{p}}}\frac{\partial f(E_{\mathbf{p}})}
{\partial E_{\mathbf{p}}}\Big].
\end{equation}
If we extend this calculation below $T_c$
we have for
$\big(\frac{\partial n}{\partial \mu}\big)_{T < T_c}$ the following expression
\begin{eqnarray}
I_1
+\frac{\Big[\sum_{\mathbf{p}}\frac{\xi_{\mathbf{p}}}{E^2_{\mathbf{p}}}\big(\frac{1-2f(E_{\mathbf{p}})}{E_{\mathbf{p}}}
+2\frac{\partial f(E_{\mathbf{p}})}{\partial E_{\mathbf{p}}}\big)\Big]^2}
{\sum_{\mathbf{p}}\frac{1}{E^2_{\mathbf{p}}}\big(\frac{1-2f(E_{\mathbf{p}})}
{E_{\mathbf{p}}}+2\frac{\partial f(E_{\mathbf{p}})}{\partial E_{\mathbf{p}}}\big)}.
\label{eq:24}
\end{eqnarray}
Note that below $T_c$ this has a similar structure as that in BCS theory.

The resulting curves are plotted as black lines in Figure~\ref{fig:1},
showing the behavior of the compressibility
both above and below $T_c$ for three
different values of the scattering length. This figure indicates
the changes as one varies from the BCS to the
BEC side of unitarity.
%
%
In this lowest order approximation which is based on the net effective
gap $\Delta$ there is no discontinuity in $\kappa$
which should be present at a phase transition.
Such a signature only arises
\cite{Chen4} when one introduces a
clearer distinction between the condensed and non-condensed
pairs.


A more physical result can be obtained using
a variant on this theory in which
the non-condensed
pairs have a finite lifetime $\gamma$ so that
\begin{equation}\label{SEpg}
\Sigma_{\textrm{pg}} = \frac {\Delta_{\textrm{pg}}^2}
{\omega + \epsilon_{\bf k} + i \gamma}.
\end{equation}
Such a broadened BCS-like self energy (with $\gamma \neq 0$)
has been applied extensively in cold gas studies, particularly
in analyzing experiments involving radio-frequency spectroscopy \cite{RFlong,Jin6}.
With this self energy
and the addition of the condensate self energy the spectral function $A(\omega,\mathbf{k})$
can be readily calculated \cite{RFlong}.
We use $n=\sum_{\mathbf{k}}\int \frac{d\omega}{\pi}A(\omega,\mathbf{k})f(\omega)$ to directly evaluate \cite{Note1}
the two contributions to the normal state
compressibility via
$\Big(\frac{\p n}{\p\mu}\Big)_{\Delta}$ and
$\Big(\frac{\p n}{\p\Delta}\Big)_{\mu}$.
As in RF experiments \cite{RFlong}, in the numerics, we take $\gamma/E_F=\alpha T/T_c$
with the constant $\alpha$ of order unity,
(although the behavior is extremely insensitive to this parameter).
Importantly, the
presence of $\gamma$ then leads to a thermodynamic
feature at $T_c$ in the spectral function. This same
thermodynamical feature is then mirrored in
the compressibility.

The results of this simple modification of the black lines
in Figure~\ref{fig:1} is plotted as red curves in the figure, which show a
discontinuity in the compressibility at $T_c$.
In the normal phase there are two competing terms which
one can see directly from comparing the first and
second terms in Eq.~(\ref{eq:18}).
The first term
leads to a contribution to the
compressibility which \textit{decreases} with decreasing
temperature above $T_c$. The second term leads to
a component which \textit{increases} with decreasing temperature
above $T_c$.
We may interpret the first (and second) of these as associated with the
fermionic (and bosonic) degrees of freedom. The growth of a fermionic gap
tends to depress $\kappa$.
At the same time the onset of bosonic degrees of freedom leads to
a large (but in contrast to Ref.~\cite{Lucheroni},) non-divergent compressibility.

Below $T_c$, on the other hand, except for a pairing gap which is no
longer the same as the order parameter, the behavior of the
compressibility is rather similar to that of BCS theory; the
decrease with decreasing $T$ reflects
the fermionic gap \cite{Note3}.
As one approaches the BEC limit the behavior becomes
progressively more temperature independent, essentially
because the fermionic parameters, such as $\Delta$
and $\mu$,
reflected in $\partial n /\partial \mu$, are
insensitive to $T$ due to the strong binding energy.
In this way, our theory provides predictions for the behavior of $\kappa$ in BCS-BEC crossover.

We emphasize that while we could, we have not adjusted the parameter $\gamma$, taken from
RF studies,
to fit experiment \cite{Zwierlein12} or Monte Carlo simulations \cite{MC12}.
Nevertheless, we point out that the step discontinuity and general
behavior for the compressibility found here is within roughly a factor of two as observed experimentally (Figure 2A of Ref.~\cite{Zwierlein12}).


\textit{Conclusions}
Past work in the literature has addressed the compressibility
\cite{StrinatiPRL12,Haussmann12} using a response function approach.
Potentially more reliable
\cite{Mahanbook,Singwi} are
thermodynamical approaches to $\kappa$. Because the compressibility
is a central means of identifying the transition, it is
of particular interest
as we do here, to find and apply a thermodynamic methodology
which does not predict an unphysical first order transition \cite{firstordertransitionpapers_full}.
An additional advantage is that a thermodynamic approach
avoids the complexity of gauge invariance and collective modes which
must be included in the superfluid state response functions, thereby
allowing
$\kappa$ to be addressed on both sides of $T_c$.

Our thermodynamic scheme has been applied on both sides of resonance,
and thus provides predictions for the BEC regime, showing a step
discontinuity which decreases from BCS to BEC, as found earlier
\cite{Chen4} for the specific heat.
We find no divergences in $\kappa$, which are avoided
\cite{Lucheroni,StrinatiPRL12}
by resumming classes of diagrams in a response function approach.
As we have emphasized here and elsewhere \cite{OurComment,TheirComment}
such resummations
have to be implemented so as not
to violate either or both the gauge invariant and $Q$-limit
Ward identities.
With improved approximations, it may be possible to recover consistency
with these three important (sum rule) constraints,
with which our full $t$-matrix
approach to BCS-BEC crossover is manifestly
compatible.

KL and YH contribution supported by NSF-MRSEC Grant
0820054. Hao Guo thanks the support by National Natural Science Foundation of China (Grants No. 11204032) and Natural Science Foundation of Jiangsu Province, China (SBK201241926).
C. C. C. acknowledges the support of the U.S. Department of Energy through the LANL/LDRD Program.


\begin{thebibliography}{29}
\bibitem[{ord()}]{order_note}
\bibinfo{note}{When constraining the form of $\Gamma^{0}$, the Ward identity
  and the $Q$-limit Ward identity are associated with two different orders of
  limits for the vanishing of $\mathbf{q}$ and $\Omega$.}
\end{thebibliography}

\end{document}